\documentclass[preprint]{aastex62}

\usepackage{graphicx}        
\usepackage{amssymb}         
\usepackage{xcolor}
\definecolor{darkblue}{rgb}{0,0,0.5}
\usepackage{epstopdf}

\DeclareFontFamily{OT1}{pzc}{}
\DeclareFontShape{OT1}{pzc}{m}{it}%
             {<-> s * [1.1500] pzcmi7t}{}
\DeclareMathAlphabet{\mathscr}{OT1}{pzc}%
                                 {m}{it}

\usepackage{amsmath}

\makeatletter
\def\mathcolor#1#{\@mathcolor{#1}}
\def\@mathcolor#1#2#3{%
  \protect\leavevmode
  \begingroup
    \color#1{#2}#3%
  \endgroup
}
\makeatother

\newcommand{\fract}[2]{\leavevmode\kern.1em
          \raise.5ex\hbox{\the\scriptfont0 #1}\kern-.1em
    \raise.15ex\hbox{\the\scriptfont0 /}\kern-.08em\lower.25ex\hbox{\the\scriptfont0 #2}}

\renewcommand{\k}{\mathbf{k}}

\newcommand{\B}{{\mathbf{B}}}

\allowdisplaybreaks[1]

\begin{document}


\title{Smoothing of MHD Shocks in Mode Conversion}

\author{Jamon D.~Pennicott}
\email{jamon.pennicott@monash.edu}
 
\author{Paul S.~Cally}
  \email{paul.cally@monash.edu}
  
\affil{School of Mathematics and Monash Centre for Astrophysics,\\ Monash University, Clayton, Victoria 3800, Australia}

\shortauthors{Pennicott \& Cally}

\shorttitle{Smoothing of MHD Shocks in Mode Conversion}

\begin{abstract}\noindent
Shock waves are simulated passing through the Alfv\'en-acoustic equipartition layer in a stratified isothermal magneto-atmosphere. The recent ray-theoretic calculations of \cite{Nun19aa} predicted smoothing of the shock through this layer, causing both the fast and slow components to emerge as continuous waves. However, it is found that the partial mode conversion expected from linear theory for oblique incidence of the shock on the magnetic field is accompanied by a smoothing of the slow-shock only, whilst the fast-shock persists. Explanations are presented based on MHD mode conversion and shock theory.
\end{abstract}



\section{Introduction}\label{sec:intro}
Magneto\-hydro\-dynamic (MHD) waves may change nature -- between fast, slow and Alfv\'en -- several times in moving through stellar atmospheres. Both fast-slow \citep{SchCal06aa} and fast-Alfv\'en \citep{CalGoo08aa,CalHan11aa} conversions have been explored in detail for linear waves \citep[reviewed by][]{CalMorRaj16aa}. Linear fast-slow conversion occurs where the Alfv\'en speed $a$ matches the sound speed $c$, and fast-to-Alfv\'en conversion is affected near where the fast wave reflects in a stratified atmosphere (provided the wave vector is not in the same vertical plane as the magnetic field lines). However, mode conversion by these processes is less well understood for nonlinear waves, in particular shocks.

Recently, \cite{Nun19aa} presented an analysis based on the generalized ray theory of \cite{Cal06aa} and \cite{SchCal06aa} (see also \citealt{TraKauBri03aa}) to argue that magnetoacoustic shocks passing through a layer where $a$ and $c$ coincide (the Alfv\'en-acoustic equipartition) not only split into fast and slow components as suggested by linear theory, but also that both the resulting fast and slow shocks are smoothed in the process. 

N\'u\~nez's analysis proceeds by examining the convergence of the Fourier components of the incoming wave. A shock discontinuity is associated with an $\mathcal{O}(1/n)$ asymptotic behaviour in wavenumber $n$ space. Faster convergence results in continuous (smooth) solutions. However, the argument is predicated on an assumption which is merely stated and not justified:
\emph{`Assuming that the passage through the conversion zone acts linearly in the waves, which is reasonable since they spend very little time there, we obtain the surprising result that the shocks which may be present in the incident wave are smoothed out in the outgoing ones, except for the single case of a pure transfer from a fast to a slow wave or vice versa.'} \citep[Conclusions]{Nun19aa}.

The purpose of this letter is to present the results of 1.5-dimensional (1.5D) simulations testing this assumption and the resulting finding that shocks are smoothed. We conclude that the assumption is in fact not valid -- the mode conversion region is advected with the slow `shock' for several seconds thereby extending the time over which conversion is operating -- and that the smoothing is only partially realized; the slow shock is smoothed but the fast shock is not. Elementary explanations for this are presented in the discussion. These suggest that the smoothing of the slow shock but persistence of the fast shock may happen even without advection of the $a=c$ equipartition layer.


\section{Model}\label{sec:model}
We adopt the simplest suitable model to explore the phenomenon: an isothermal gravitationally stratified 1.5D atmosphere with uniform magnetic field $\B=B_0(\sin\theta,0,\cos\theta)$ inclined at angle $\theta$ to the vertical. The magnetic field is chosen such that the Alfv\'en speed $a$ equals the sound speed $c$ at a sufficient altitude that a fast (i.e. acoustic) wave injected at the bottom of the computational domain has enough space to shock before reaching the equipartition level $a=c$.

The computational box stretches $2.0$ Mm both above and below the $z=0$ height.  The equilibrium density scale height is $h=175.731$ km, the uniform sound speed is $c=8.958$ $\text{km}\,\text{s}^{-1}$, the adiabatic index is $\gamma=5/3$, and the magnetic field strength is $B_{0}=0.9$ kG oriented at angle $\theta$ to the vertical. The $a=c$ equipartition level is situated at height $z_0=125$ km. The acoustic cutoff frequency is $\omega_c=c/2h$, or 4.0 mHz. There is nothing particularly special about these values, though they are broadly characteristic of the active solar chromosphere. Changing $B$ simply moves the equipartition level up or down. Similar results may be expected for other stellar atmospheric models.

The numerical code used to solve the non-linear MHD equations is Lare2d \citep{ArbLonGer01aa}.  The grid is comprised of 8192 cells in the vertical direction, giving a resolution of $0.488$ km.

A shock viscosity term is employed to avoid the Gibbs phenomenon whilst not artificially smoothing the shock front. It can be seen that the shock fronts remain sharp where predicted throughout the simulations.  An artificial cooling term is applied to avoid thermal runaway from shock heating, allowing the atmosphere to remain close to its initial profile.  This term has the form of an exponentially weighted moving average, where the degree of weighting is set at $\alpha_w = 0.05$. When the cooling term is removed, the results show a $<1\%$ difference within a scale height either side of the $a=c$ layer, our main area of interest.

Exact solutions for mode conversion and transmission are known for the linear model in terms of ${}_2F_3$ generalized hyper\-geo\-metric functions \citep{HanCalDon16aa}. Transmission decreases with increasing attack angle between the wave vector and the magnetic field \citep{SchCal06aa}. According to \cite{Nun19aa}, an acoustic shock with zero attack angle should pass through $a=c$ unchanged. On the other hand, if we orient the magnetic field away from vertical, substantial mode splitting will occur and the resulting fast and slow `shocks' should be smoothed throughout this process if N\'u\~nez's analysis is correct.

The fast wave above $a=c$ is predominantly magnetic in nature, and will not reflect due to the ever-increasing Alfv\'en speed gradient as the wave is exactly vertical. In any case, that is beyond the phenomenon of interest here. The acoustic slow wave also propagates upward unhindered and becomes progressively more aligned with the magnetic field as the plasma beta reduces towards zero, whilst the fast wave becomes ever more transverse with height. This provides a convenient way to distinguish them in $a\gg c$, as well as via their propagation speeds. 

Specifically, we plot the transverse and parallel-to-the-field plasma velocities $v_\perp$ and $v_\parallel$, as well as the density $\rho$ scaled by the equilibrium density $\rho_0(z)$. It is important to note that these `$\perp$' and `$\parallel$' components are with respect to the actual (perturbed) magnetic field, not the original field. Since the shock front compresses the plasma vertically by up to a factor of 4, and the field lines are tied to the plasma, the perturbed field inclination is greatly increased.


\section{Results}\label{sec:results}
A wave driver is placed at the base $z=-2.0$ Mm, and delivers an angular-frequency-$\omega_0$ half-period sinusoidal vertical velocity burst of amplitude $V_0$. The driver creates a strictly vertical wave vector. The plasma velocity is also initially vertical, until the inclined magnetic field takes effect as the plasma beta decreases with increasing altitude. For concreteness, we adopt $\omega_0=2\omega_c = 8.0$ mHz. This corresponds to a broad frequency spectrum 
\begin{equation}\label{F_fourier}
   \tilde F(f) = -f_0 \frac{\cos(\pi f/2f_0)}{(f^2-f_0^2)\pi}
\end{equation}
centred at zero, where $f=\omega/2\pi$. However, the acoustic cutoff only lets frequencies above 4 mHz propagate upward. This is confirmed numerically. Using drivers lasting several half-periods allows us to localize the driving frequency more precisely, but introduces other numerical complications, particularly first shocks reflect off the top of the computational domain and complicate behaviour in the region of interest. For cleanest results, the half-period driver has been found to be best. In any case, it makes very clean shocks from a broad spectrum of frequencies above $f_c$.

\citet{KalRosBod94aa} have discussed the evolution of linear and nonlinear waves in non-magnetic 1D isothermal atmospheres. The wakes produced by the initial pulse follow each other by typically one cut-off period, supporting the theory that the atmosphere rings at the cut-off frequency. We restrict our results to tracking the initial pulse produced by the 8.0 mHz driver.

\begin{figure*}[htbp]
    \centering
    \includegraphics[width=.9\textwidth]{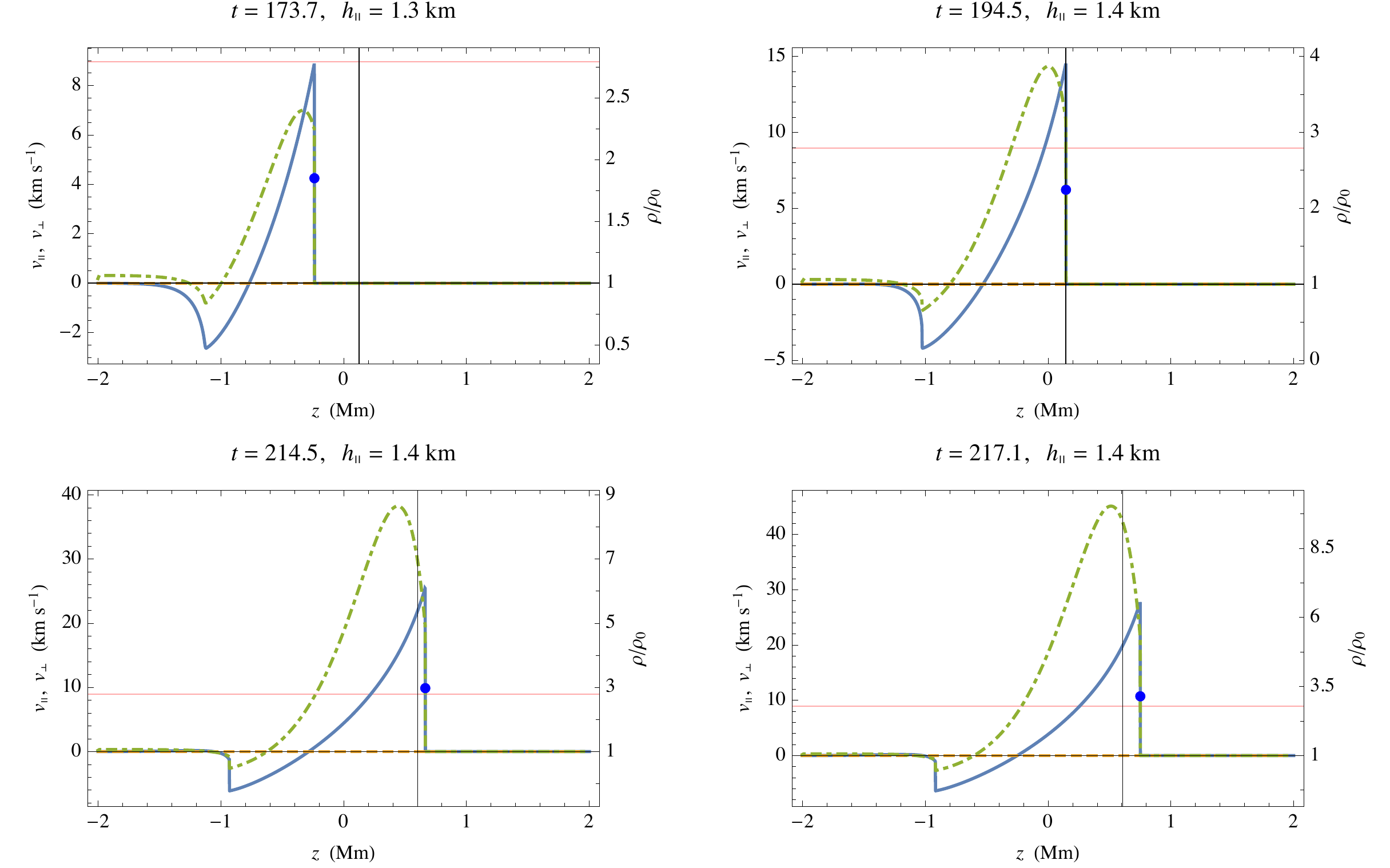}
    \caption{$\theta=0^\circ$: The blue solid line, orange dashed line (zero in this case) and green dot-dashed line correspond to $v_{\parallel}$, $v_\bot$ and $\rho/\rho_{0}$ respectively. The shock front is formed and travels through the $a=c$ layer unaltered.  The solid vertical line is where $a=c$, which moves in response to the incident wave. The pink horizontal line is the equilibrium sound speed. The blue dot indicates the steepest point in the acoustic shock. The top-right panel indicates that the shock reaches the conversion layer with almost-maximal density contrast ($X=4$).}
    \label{fig:theta0}
\end{figure*}

The velocity perturbation increases exponentially with height until a shock is formed well below the $a=c$ layer. When the magnetic field is vertical, the shock front remains sharp throughout the entire simulation and travels unhindered through to the top of the computational box, as expected (Fig.~\ref{fig:theta0}). Increasing $\theta$, the angle of the magnetic field from vertical, produces a splitting of the wave into its slow and fast components around the $a=c$ layer. When $a>c$, the fast wave accelerates out in front of the slow wave and the two wave types can be easily distinguished. 

We choose to look separately at velocities both parallel and transverse to the (actual) magnetic field. This separation allows an approximation of the two wave types, however it is not precise. This can be seen by the fast wave appearing in the parallel velocity for $\theta>0$, and vice versa for the slow wave.

We also look at the in-place compression ratio $\rho/\rho_{0}$, where $\rho$ is the density value at the given time and $\rho_{0}$ is the initial background density value at the same position. This is not the same as the classical shock-jump compression ratio $X=\rho_2/\rho_1$ from the pre-shock to post-shock regions (`1' and `2' respectively), though they do correspond immediately after a shock front passes. For an oblique compressive shock, where the magnetic field is not completely perpendicular to the shock front, the limiting compression ratio for $\gamma=5/3$ is 4 \citep{Pri82aa}. For these simulations, this remains true across the shock front as expected; however the values behind the front can exceed this ratio due to the transport of fluid via the propagating shock. Deeper, higher density fluid is carried upwards and is then related to the original density values of greater heights giving the appearance of a much higher compression ratio. To eliminate this, the movement of each fluid particle would need to be traced throughout the simulation and the densities compared to those at their original equilibrium height, which is an unnecessary complication.

\begin{figure*}[htbp]
    \centering
    \includegraphics[width=.9\textwidth]{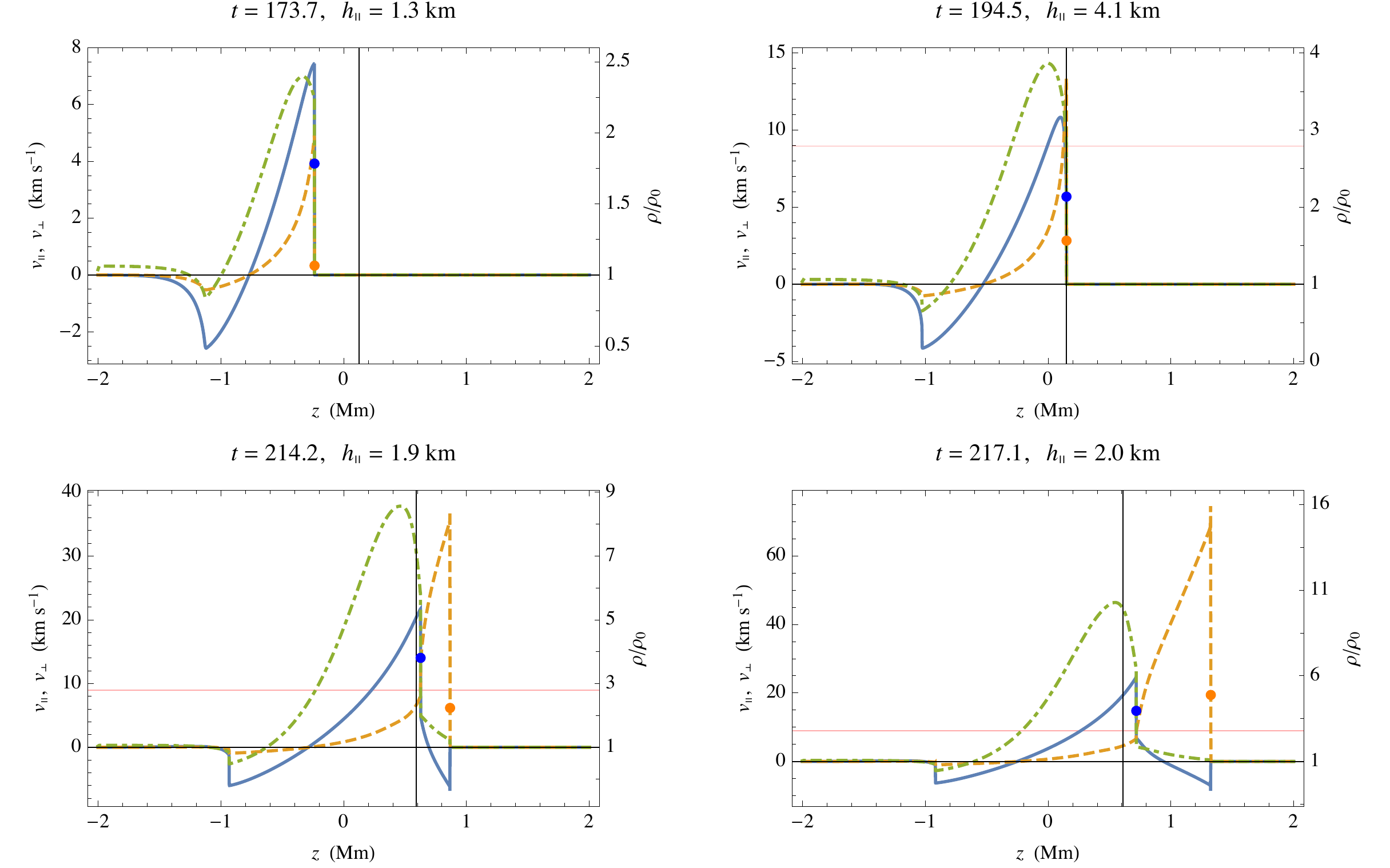}
    \caption{$\theta=15^\circ$: At the $a=c$ layer, the slow shock is smoothed slightly, before steepening again to produce another shock front. The fast shock remains sharp and propagates unaltered. The orange dot shows the location of the steepest point in the fast shock.}
    \label{fig:theta15}
\end{figure*}

A non-zero attack angle gives rise to the splitting of the fast and slow components around $a=c$. However both the fast and slow shock remain sharp in the simulations when $\theta$ is less than about $15^\circ$. (In reality, we believe that the slow shock will smooth for any non-zero attack angle, but this is computationally imperceptible at these very narrow incidences.) After this point, we begin to see smoothing of the slow component as it separates from the fast shock and traverses the layers around $a=c$. As the slow wave exits the $a=c$ layer, it begins to steepen again to `re-shock', best seen in Fig.~\ref{fig:theta15}.  Conversely, the fast shock remains sharp throughout and continues to propagate through the atmosphere unaltered in its form. It does not reflect because it is vertical.

\begin{figure*}[htbp]
    \centering
    \includegraphics[width=.9\textwidth]{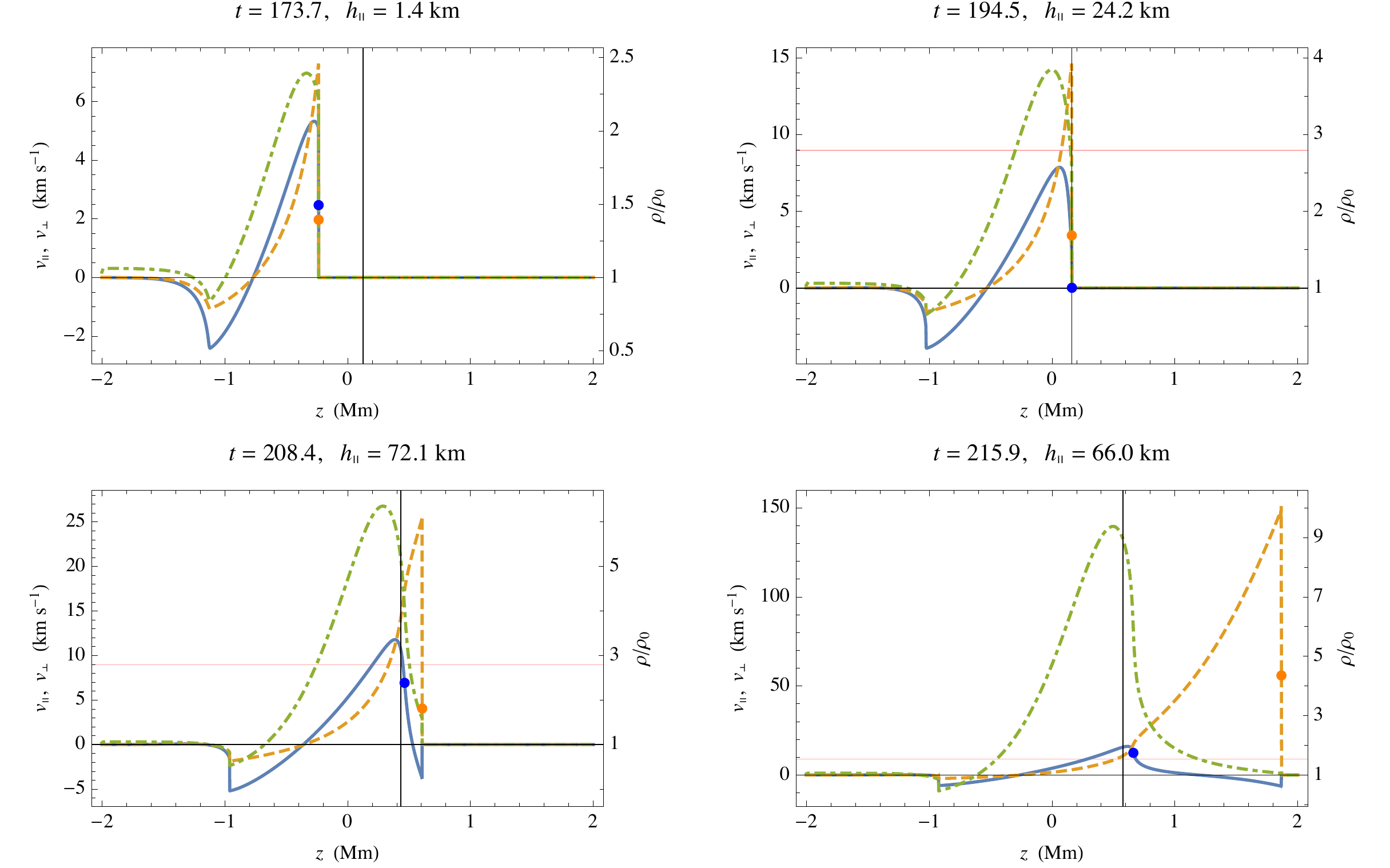}
    \caption{$\theta=30^\circ$: At this stage, the increased magnetic field inclination produces an easily identifiable smoothing of the slow shock around $a=c$ before beginning to steepen again.  Again, the fast shock remains sharp and propagates freely.}
    \label{fig:theta30}
\end{figure*}

\begin{figure*}[htbp]
    \centering
    \includegraphics[width=.9\textwidth]{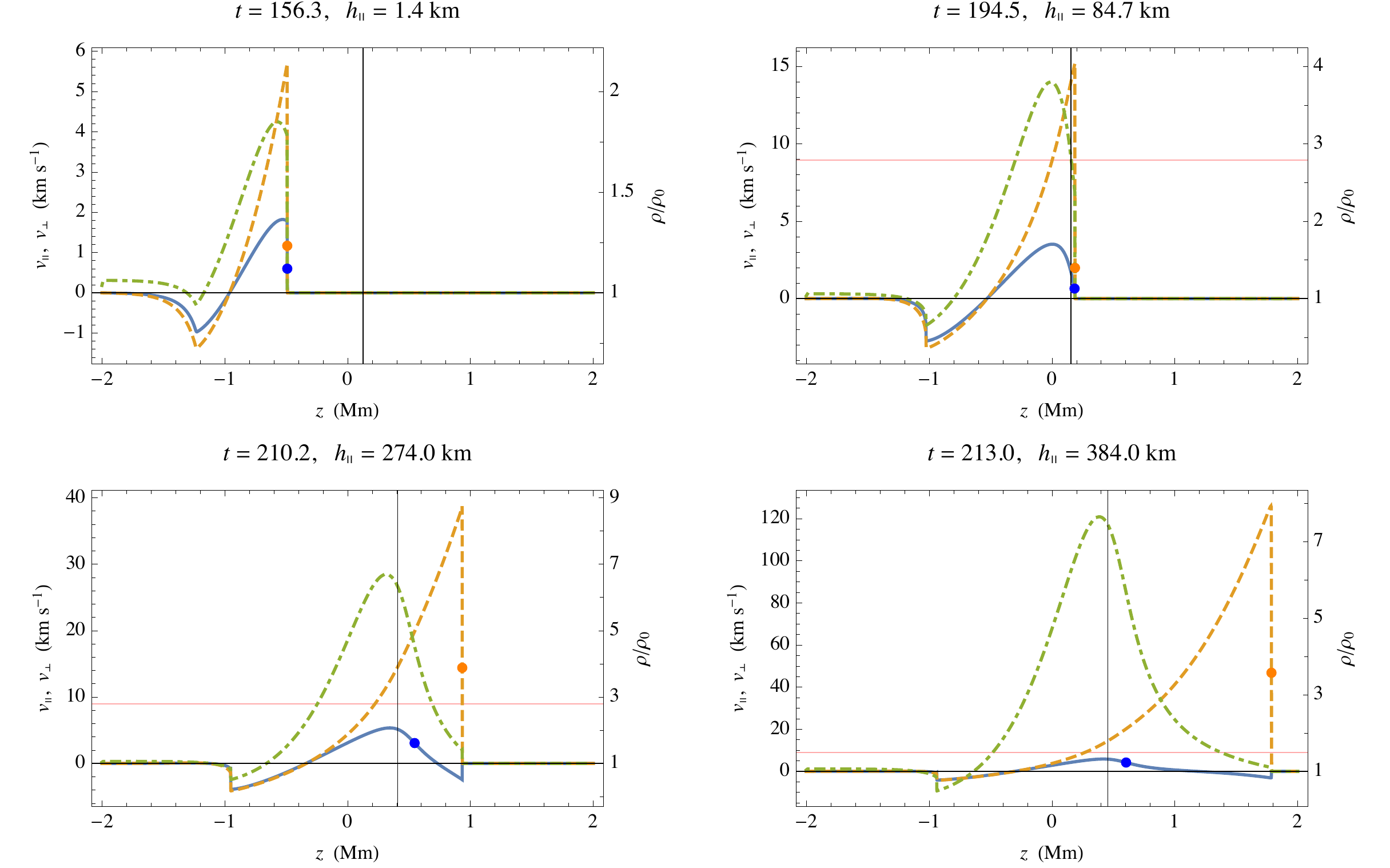}
    \caption{$\theta=60^\circ$: Heavy smoothing of the slow shock through $a=c$.  The fast shock remains sharp and propagates freely again. The transmitted slow wave is subsonic and much too weak to re-shock before the fast shock reaches the top of the computational box.}
    \label{fig:theta60}
\end{figure*}

 Figures \ref{fig:theta15}, \ref{fig:theta30}, and \ref{fig:theta60} display the evolution of the $\theta=15^\circ$, $30^\circ$, and $60^\circ$ cases respectively. The greater the value of $\theta$, the more smoothing of the slow component is seen.  We quantify this by measuring the steepness of the slow shock at its inflection point. The blue dots shown on the `shock' front in each case is where $v_{\parallel}$ has its inflection point, i.e., where the `shock' is at its steepest.  The value of $h_{\parallel}$ given above each panel is the `velocity scale height' there and is defined as
\begin{equation}
h_{\parallel}=\left|v_{\parallel,\text{max}}\middle/\left(\frac{dv_{\parallel}}{dz}\right)\right|,
\end{equation}
which is a measure of shock steepness. 
A small $h_{\parallel}$ corresponds to a steep shock front (numerically limited).

\begin{figure*}[htbp]
    \centering
    \includegraphics[width=.5\textwidth]{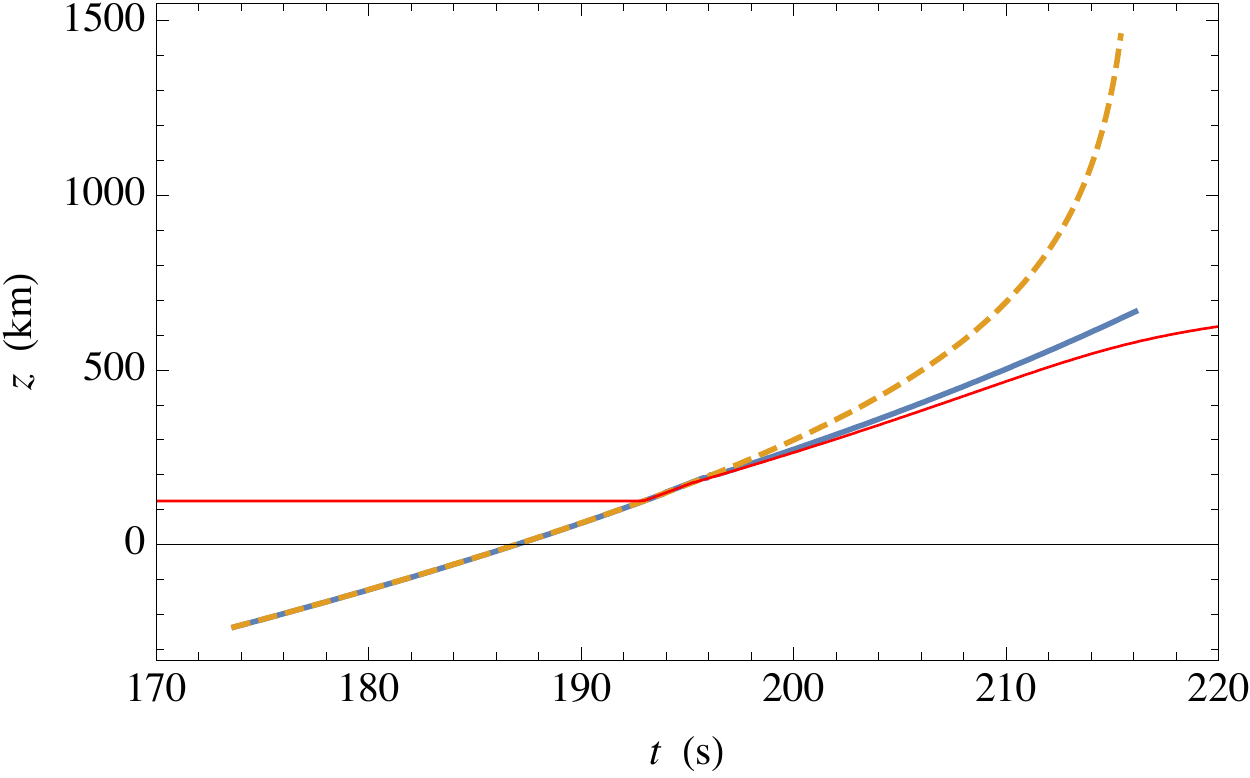}
    \caption{Heights in time of the `shock' seen in $v_\parallel$ (blue curve; the shock is actually smoothed for a while after meeting the equipartion level) and the shock in $v_\perp$ (dashed orange curve) as well as the position of the $a=c$ equipartition level (red), for the case of the $30^\circ$ inclined magnetic field. Clearly, the equipartition level is advected with the slow shock for several seconds.}
    \label{fig:cross30}
\end{figure*}

\cite{Nun19aa} assumes the passage through the conversion zone acts linearly on the waves, which is based on the supposition that they spend very little time in that area. However, the incoming shock front acts to drag the $a=c$ layer along with it (Fig.~\ref{fig:cross30}), which can increase in height by up to 400 km for low $\theta$.  The shock front itself can spend multiple seconds within close proximity of the $a=c$ layer.

\section{Discussion and Conclusions}
1.5D simulations were conducted using a bottom-driven 8.0 mHz half-period pulse injected into an isothermal, gravitationally stratified atmosphere with the angle the magnetic field makes from vertical being varied from $\theta = 0^{\circ}$ to $60^{\circ}$. Only power above the acoustic cutoff of 4 mHz propagates upward. The perturbation was allowed to freely evolve, forming a shock well below the Alfv\'en-acoustic equipartition ($a=c$) layer. If the attack angle $\alpha$ between the (vertical) wave vector $\mathbf{k}$ and magnetic field is non-zero, the shock splits into its slow and fast components around $a=c$. For $\theta\gtrsim 15^{\circ}$, the slow-shock in our simulations shows smoothing through this area, then begins to re-shock upon exiting.  The fast-shock, however, continues to propagate freely throughout the atmosphere regardless of the value of $\theta$.  This is in only partial agreement with \cite{Nun19aa}, who stated that both the fast and slow components of the shock should be smoothed. A possible reason for the discrepancy is that the $a=c$ equipartition region is advected with the shock for several seconds, making the interaction intrinsically nonlinear.

\begin{figure*}[htbp]
    \centering
    \includegraphics[width=0.5\textwidth]{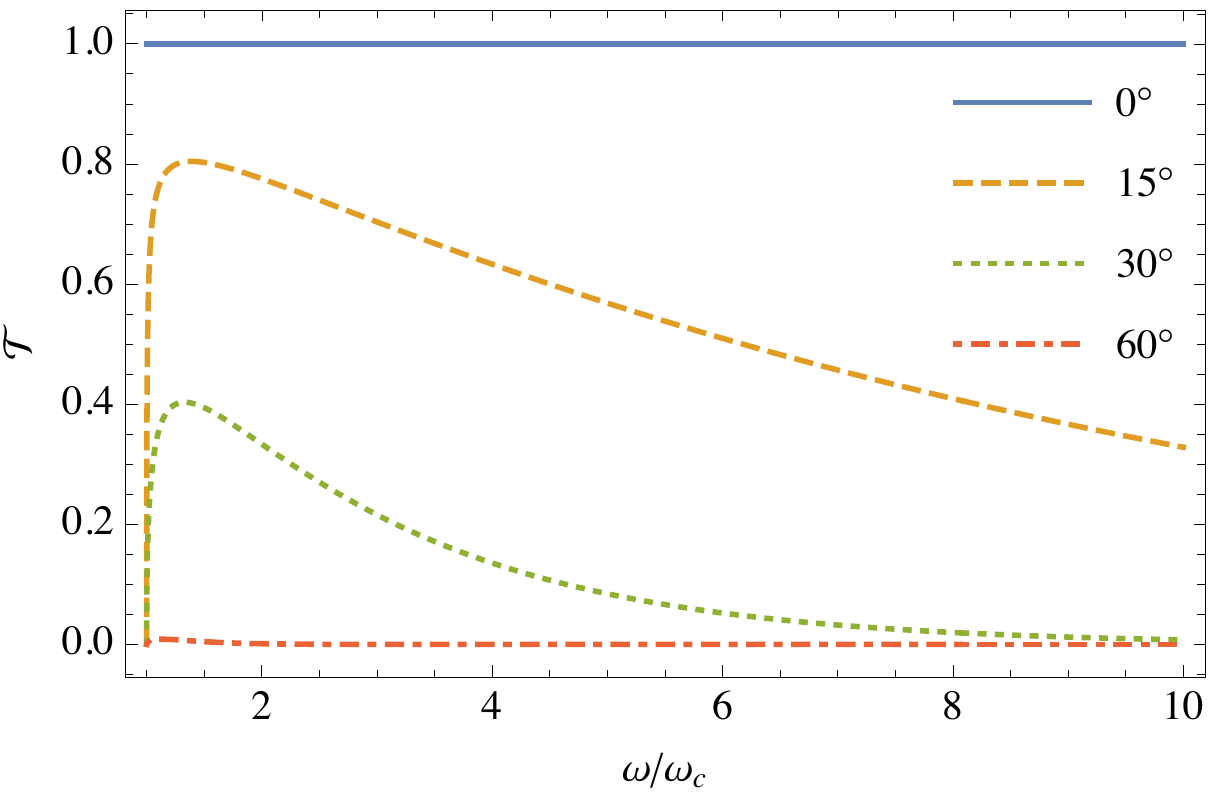}
    \caption{Exact linear transmission coefficients for vertical waves in an isothermal atmosphere as function of frequency measured relative to the acoustic cutoff frequency, for magnetic field inclinations $\theta=0^\circ$, $15^\circ$, $30^\circ$, and $60^\circ$ (top to bottom). Adiabatic index $\gamma=5/3$ is assumed.}
    \label{fig:vert_T}
\end{figure*}

However, it is not clear that N\'u\~nez's result is correct even without this effect. Figure \ref{fig:vert_T} shows the exact linear transmission coefficient $\mathcal{T}$ for a vertical wave in an isothermal atmosphere \citep{Cal09aa} as a function of frequency with the four field inclinations discussed above. The complement, $\mathcal{C}=1-\mathcal{T}$ is the conversion coefficient, i.e., the fraction of energy going into the fast wave above $a=c$. Of course, it starts at the cutoff frequency $\omega_c$. Any shock wave discontinuity is necessarily constructed from high-order Fourier modes with amplitudes asymptotically proportional to the inverse of the vertical wavenumber $k_z$. In turn, $k_z$ is proportional to $\omega$ through a dispersion relation. Figure \ref{fig:vert_T} shows that, especially at greater magnetic field inclinations, high-frequency high-wavenumber modes are not transmitted to any appreciable degree, thereby smoothing the incident acoustic shock. On the other hand, these small-scale Fourier modes are essentially perfectly \emph{converted} to fast waves, which are therefore sharp. This is in accord with our simulations. The advection of the $a=c$ level with the shock could only amplify this effect.

The same conclusion follows from the approximate WKB-based formula for transmission \citep{SchCal06aa},
\begin{equation}
    \mathcal{T}\approx
    \exp\left[-\pi h_s |\k| \sin^2\alpha\right]_{a=c},
\end{equation}
where $h_s=[d(a^2/c^2)/ds]_{a=c}^{-1}$ is the thickness of the conversion layer along the direction of the wave vector $\k$, and $\alpha$ is the attack angle between $\k$ and $\B$. In our case $\alpha=\theta$, $s=z$, and $\mathcal{T}\to0$ exponentially with increasing $|\k|$, faster at greater $\theta$. Again, this is in accord with our findings. The advection of $a=c$ with the shock may effectively increase $h_s$ and thereby decrease $\mathcal{T}$ and increase $\mathcal{C}$, as suggested above.

A further perspective is afforded by the jump relations for MHD shocks in uniform media, and in particular the so-called \emph{shock adiabatic}, a bi-cubic polynomial for the shock speed $v_1$ dependent on adiabatic index $\gamma$, magnetic field inclination $\theta_1$ in the pre-shock plasma, shock compression ratio $X=\rho_2/\rho_1$ (where $1<X<(\gamma+1)/(\gamma-1)$, as for hydrodynamic shocks), and pre-shock sound and Alfv\'en speeds $c_1$ and $a_1$ respectively \citep{Fit14aa}:
\begin{equation}
\begin{split}
    A(v_1,\theta_1,\gamma,c_1,a_1)=&(v_1^2- X \cos^2\theta_1 a_1^2)^2  \left\{\left[(\gamma+1)-(\gamma-1)X\right]v_1^2-2X c_1^2\right\}\\
   & -X\sin^2\theta_1 a_1^2\left\{\left[\gamma+(2-\gamma)X\right]-\left[(\gamma+1)-(\gamma-1)X\right]X \cos^2\theta_1 a_1^2\right\}=0.
\end{split}
\end{equation}
The post-shock region is denoted by label `2'. For $\gamma=5/3$, compression is limited to $1<X<4$. This should give a good local description of shocks in a stratified atmosphere.

\begin{figure*}[htbp]
    \centering
    \includegraphics[width=.5\textwidth]{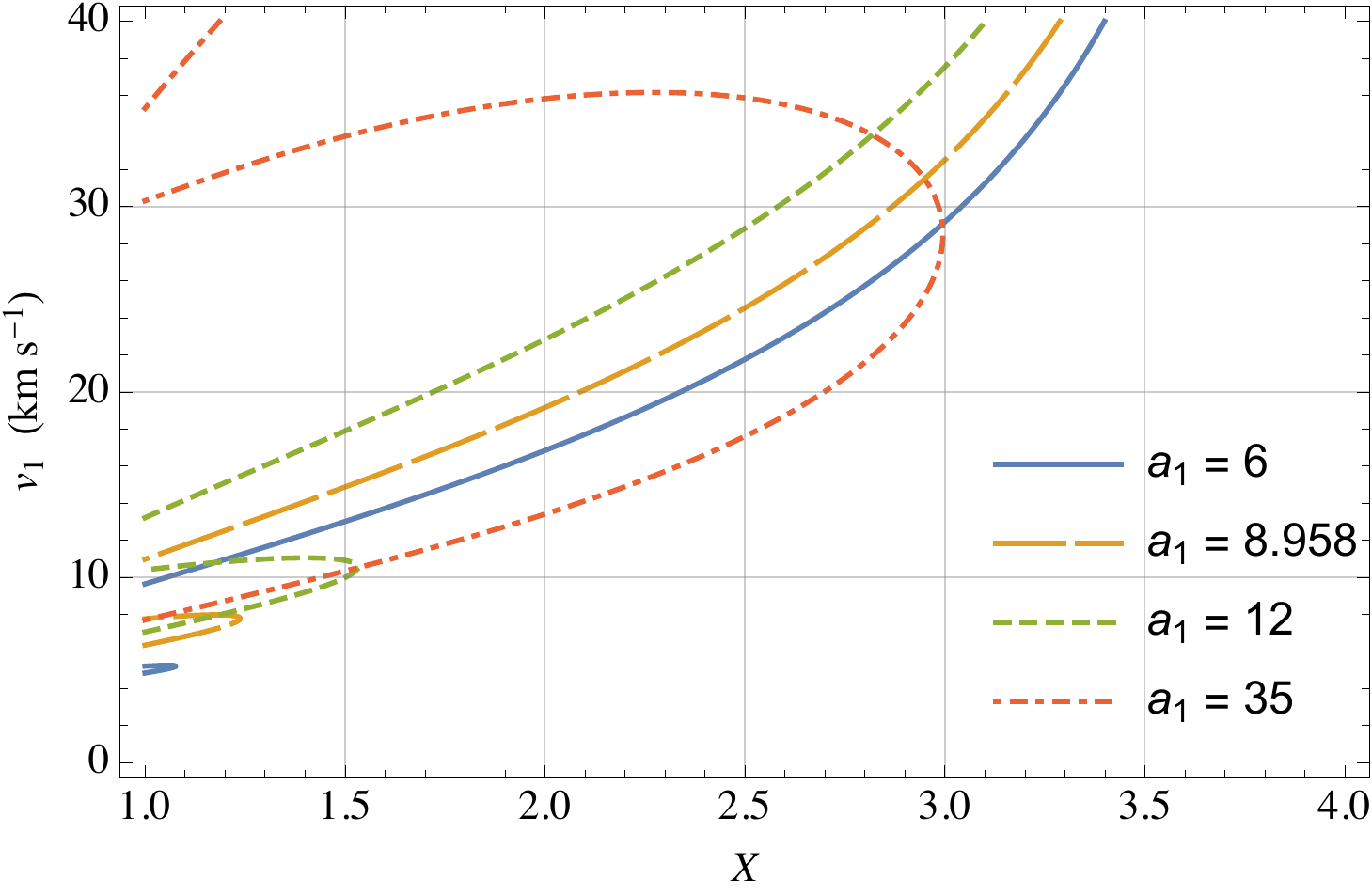}
    \caption{Adiabats as functions of compression ratio $X$ and shock speed $v_1$ for $\gamma=5/3$, $\theta_1=30^\circ$, $c_1=8.958$ km $\rm s^{-1}$, and four values of $a_1$: 6 km~$\rm s^{-1}$ (full blue), $c_1=8.958$ km~$\rm s^{-1}$ (long-dashed orange), 12 km~$\rm s^{-1}$ (short-dashed green), and 35 km~$\rm s^{-1}$ (dot-dashed red).}
    \label{fig:adiabat30}
\end{figure*}

Consider the case $\gamma=5/3$, $\theta=30^\circ$, corresponding to Fig.~\ref{fig:theta30}. Adiabats for our canonical sound speed $c_1=8.958$ km~$\rm s^{-1}$ and four Alfv\'en speeds are plotted in Fig.~\ref{fig:adiabat30}. For each Alfv\'en speed case, the upper curve corresponds to the fast shock. The loops beneath correspond to the slow shock (lower branch) and the intermediate or Alfv\'en shock (upper branch). The intermediate shock is not excited in our simulations. Clearly, the fast shock exists for each $1<X<4$, but the slow shock is absent if $X$ is to the right of its loop. As $a_1$ increases with height in the stratified atmosphere, the intermediate/fast loop extends to greater $X$ and the shock redevelops. For example, at the height where $a_1=12$, a slow shock may only exist for $X\lesssim1.53$, but for $a_1=35$ there is a slow shock for all $X\lesssim2.99$.

This is consistent with our simulations (Fig.~\ref{fig:theta30}), in that the compression is large across the slow wave front in the bottom two panels, where the somewhat-steep slow wave has passed beyond $a=c$. This means there is no available slow shock solution, providing another explanation for its absence in the simulations.

On the other hand, the fast shock always has a solution. Because $X$ is close to 1 (and gets closer with height) across the fast shock in the bottom panels, the fast shock travels at a speed only just above the Alfv\'en speed $a_1$.

The reason for the slow wave above $a=c$ inheriting most of the density contrast is difficult to see from the exact mathematics, but it is plausible conceptually. The slow wave above the equipartition height is the primarily-acoustic successor to the incident \emph{fast} shock below $a=c$, which is also mainly acoustic. It is therefore unsurprising that density is largely tied to these waves. On the other hand, the fast wave in $a>c$ is more weakly coupled to density, and in fact becomes incompressive in the $a\gg c$ limit. Irrespective of these broad observations, the numerical solutions do suggest that the slow wave in $a>c$ accounts for most of the density contrast, and therefore cannot shock until $a$ increases sufficiently for a shock solution to become available (see the 35 km~$\rm s^{-1}$ lobe in Fig.~\ref{fig:adiabat30}).

Implications for solar and stellar chromospheres might include reduced acoustic wave heating near and above the equipartition level, both because the acoustic shock is smoothed and because energy is removed from the acoustic wave. Any enhancement of energy in the fast shock due to equipartition-level advection might benefit Alfv\'en wave production by fast/Alfv\'en conversion higher in the atmosphere.




\bibliographystyle{aasjournal}        
\bibliography{fred}

\begin{thebibliography}{}
\expandafter\ifx\csname natexlab\endcsname\relax\def\natexlab#1{#1}\fi
\providecommand{\url}[1]{\href{#1}{#1}}
\providecommand{\dodoi}[1]{doi:~\href{http://doi.org/#1}{\nolinkurl{#1}}}
\providecommand{\doeprint}[1]{\href{http://ascl.net/#1}{\nolinkurl{http://ascl.net/#1}}}
\providecommand{\doarXiv}[1]{\href{https://arxiv.org/abs/#1}{\nolinkurl{https://arxiv.org/abs/#1}}}

\bibitem[{{Arber} {et~al.}(2001){Arber}, {Longbottom}, {Gerrard}, \&
  {Milne}}]{ArbLonGer01aa}
{Arber}, T.~D., {Longbottom}, A.~W., {Gerrard}, C.~L., \& {Milne}, A.~M. 2001,
  Journal of Computational Physics, 171, 151, \dodoi{10.1006/jcph.2001.6780}

\bibitem[{{Cally}(2006)}]{Cal06aa}
{Cally}, P.~S. 2006, Royal Society of London Philosophical Transactions Series
  A, 364, 333

\bibitem[{{Cally}(2009)}]{Cal09aa}
---. 2009, \solphys, 254, 241, \dodoi{10.1007/s11207-008-9290-9}

\bibitem[{{Cally} \& {Goossens}(2008)}]{CalGoo08aa}
{Cally}, P.~S., \& {Goossens}, M. 2008, \solphys, 251, 251,
  \dodoi{10.1007/s11207-007-9086-3}

\bibitem[{{Cally} \& {Hansen}(2011)}]{CalHan11aa}
{Cally}, P.~S., \& {Hansen}, S.~C. 2011, \apj, 738, 119,
  \dodoi{10.1088/0004-637X/738/2/119}

\bibitem[{{Cally} {et~al.}(2016){Cally}, {Moradi}, \&
  {Rajaguru}}]{CalMorRaj16aa}
{Cally}, P.~S., {Moradi}, H., \& {Rajaguru}, S.~P. 2016, Washington DC American
  Geophysical Union Geophysical Monograph Series, 216, 489,
  \dodoi{10.1002/9781119055006.ch28}

\bibitem[{{Fitzpatrick}({2014})}]{Fit14aa}
{Fitzpatrick}, R. {2014}, {Plasma Physics: An Introduction} ({Boca Raton}: CRC
  Press)

\bibitem[{Hansen {et~al.}(2016)Hansen, Cally, \& Donea}]{HanCalDon16aa}
Hansen, S.~C., Cally, P.~S., \& Donea, A.-C. 2016, \mnras, 456, 1826,
  \dodoi{10.1093/mnras/stv2770}

\bibitem[{{Kalkofen} {et~al.}(1994){Kalkofen}, {Rossi}, {Bodo}, \&
  {Massaglia}}]{KalRosBod94aa}
{Kalkofen}, W., {Rossi}, P., {Bodo}, G., \& {Massaglia}, S. 1994, \aap, 284,
  976

\bibitem[{{N{\'u}{\~n}ez}(2019)}]{Nun19aa}
{N{\'u}{\~n}ez}, M. 2019, EPL (Europhysics Letters), 125, 44002,
  \dodoi{10.1209/0295-5075/125/44002}

\bibitem[{Priest(1982)}]{Pri82aa}
Priest, E.~R. 1982, Solar Magnetohydrodynamics (Dordrecht: D. Reidel)

\bibitem[{{Schunker} \& {Cally}(2006)}]{SchCal06aa}
{Schunker}, H., \& {Cally}, P.~S. 2006, \mnras, 372, 551,
  \dodoi{10.1111/j.1365-2966.2006.10855.x}

\bibitem[{{Tracy} {et~al.}(2003){Tracy}, {Kaufman}, \&
  {Brizard}}]{TraKauBri03aa}
{Tracy}, E.~R., {Kaufman}, A.~N., \& {Brizard}, A.~J. 2003, Physics of Plasmas,
  10, 2147, \dodoi{10.1063/1.1543579}

\end{thebibliography}

\end{document}